\documentclass[11pt,review,1p,number,sort&compress,a4paper]{elsarticle}

\usepackage{epsfig}
\usepackage{color,amsmath,amssymb}
\usepackage{longtable}
\usepackage{rotating}
\usepackage{lscape}
\usepackage{supertabular}
\usepackage{soul}
\usepackage{multirow}

\newcommand{\3}{$_{3}$}
\newcommand{\4}{$_{4}$}
\newcommand{\cm}{cm$^{-1}$}

\newcommand{\lnr}{^{\ell}}
\newcommand{\p}{^\prime}
\newcommand{\pp}{^{\prime\prime}}

\newcommand{\Td}{${T}_{\rm d}$(M)}

\newcommand{\ai}{\textit{ab initio}}

\newcommand{\schr}{Schr\"{o}dinger}
\newcommand{\vecr}[1]{\frac{\vec{r}_{#1}}{\vert \vec{r}_{#1}\vert}}

\journal{J. Molec. Spectrosc.}

\begin{document}

\begin{frontmatter}

\title{Vibrational transition moments of CH$_4$ from first principles}

\author{Sergei~N.~Yurchenko\corref{cr1}}
\ead{s.yurchenko@ucl.ac.uk}

\author{Jonathan Tennyson}
\ead{j.tennyson@ucl.ac.uk}

\author{Robert J. Barber}
\ead{rjb@star.ucl.ac.uk}

\address{Department of Physics and Astronomy, University College London, London, WC1E 6BT, UK}

\author{Walter Thiel}
\ead{thiel@mpi-muelheim.mpg.de}

\address{Max-Planck-Institut f{\"u}r Kohlenforschung, Kaiser-Wilhelm-Platz 1, D--45470 M{\"u}lheim~an~der~Ruhr, Germany}

\cortext[cr1]{Corresponding author}


\date{\today}

\begin{abstract}

A new nine-dimensional (9D), \textit{ab initio} electric dipole moment surface (DMS) of methane in its ground electronic state is presented. The DMS is computed using an explicitly correlated coupled cluster CCSD(T)-F12 method in conjunction with an F12-optimized correlation consistent basis set of the TZ-family. A symmetrized molecular bond representation is used to parameterise the 9D DMS in terms of sixth-order polynomials. Vibrational transition moments as well as band intensities for a large number of IR-active vibrational bands of $^{12}$CH$_4$ are computed by vibrationally averaging the \textit{ab initio} dipole moment components. The vibrational wavefunctions required for these averages are computed variationally using the program TROVE and a new `spectroscopic' $^{12}$CH$_4$ potential energy surface. The new DMS will be used to produce a hot line list for $^{12}$CH$_4$.

\end{abstract}


\end{frontmatter}


\newpage

\section{Introduction}
\label{s:intro}

Methane plays an important role in atmospheric and astrophysical chemistry. Its
rotation-vibration spectrum is of key importance for models of
the atmospheres of bodies ranging from Titan to cool stars. However
the lack of precise data on methane spectra, particularly at higher
temperatures, has severely limited models for atmospheres as
diverse as Jupiter \cite{jt198}, exoplanets  \cite{svt08,jt495} and
brown dwarfs \cite{96GeKuWi.CH4}. Any temperature-dependent model
of the methane spectrum requires a reliable treatment of the
associated transition intensities. The construction of
a reliable electric dipole moment surface (DMS) which can be used to
generate such intensities concerns us in this paper.

Theoretically, a number of \ai\ potential energy surfaces (PES) computed
at different
levels of theory have been reported
\cite{84DuHaSc.CH4,95LeMaTa.CH4,01ScPaxx.CH4,04MaQuxx.CH4,06OyYaTa.CH4,11NiReTy.CH4}.
The surface by \citet{01ScPaxx.CH4} has been used in many benchmark studies
as well as for accurate predictions of vibrational and ro-vibrational
energies of CH\4. However, to our knowledge no accurate and
comprehensive \ai\ DMSs of CH\4\ have been published so far. In
some cases \ai\ DMSs were used, but not actually provided, for example
by \citet{06OyYaTa.CH4} (at the CCSD(T)/cc-pVTZ and MP2/cc-pVTZ level of
theory) and \citet{09WaScSh.CH4} (RCCSD(T)/aug-cc-pVTZ). An empirical
DMS of limited accuracy was developed by
\citet{94HoMaQu.CH4}.

The goal of this paper is to bridge this gap and provide
an accurate and detailed \ai\ DMS of CH\4. To this end we
employed an explicitly correlated (F12) coupled cluster method in conjunction with
an F12-optimized basis set to
generate a nine-dimensional (9D) DMS of CH\4. This surface is then
used to compute vibrational transition moments and vibrational band
intensities of $^{12}$CH\4\ for a large number of IR-active
transitions in the range between 0 and 10000~\cm. This is done by
vibrationally averaging the DMS over vibrational functions of
$^{12}$CH\4\ computed through variationally solving the nuclear motion
\schr\ equation with the TROVE program \cite{07YuThJe.method}. This
program has been used successfully for accurate and extensive ro-vibrational
calculations on different tetratomic molecules
\cite{10YuCaYa.SbH3,08OvThYu.PH3.JCP,10YaYuJe.HSOH,11YaYuJe.H2CO,11YaYuRi.H2CS,11YuBaTe.NH3,13PoKoMa.SO3,13PoKoMa.HOOH}.
The present work is the first application of TROVE to a larger
system.  Other variational approaches capable of
treating this pentatomic molecule include the Lanczos-method developed
by Carrington and Wang
\cite{04WaCaxx.CH4,03WaCaxx.CH4,04WaCaxx.CH4.a,04WaCaxx.CH4.b},
MULTIMODE \cite{99CaShBo.CH4,00CaBoxx.CH4,06WuHuCa.CH4,04ChTrBo.CH4}, and GENIUSH \cite{09MaCzCs.method}
as well as the methods of \citet{97Haxxxx.CH4}, \citet{jt291},
\citet{02Scxxxx.CH4},
and \citet{09Nixxxx.CH4}.
The rigorous quantum dynamics
algorithm employed by \citet{04Yuxxxx.CH4,09Yuxxxx.CH4} should be also
mentioned as an alternative accurate approach employed for calculating
vibrational energies of CH\4.

On the experimental side, electric transition dipole moments for CH\4\ are
the subject of a large number of publications
\cite{85HiLoBr.CH4,89BrLoHi.CH4,92HiLoBr.CH4,93HiBaBr.CH4,94HiChTo.CH4,96OuHiLo.CH4,98MoAbLo.CH4,00FeChJo.CH4,01HiRoLo.CH4,01RoHiLo.CH4,09AlBaBo.CH4},
where effective dipole moment parameters were derived either through
fits to intensity measurements or using perturbation theory.  The
most recent and comprehensive collection of the effective dipole
moment parameters describing excitations in the region up to
4800 \cm\ was reported recently by \citet{09AlBaBo.CH4}. Many of
these studies also provide band intensities. We use some of these data to
validate our vibrational transition moments as well as our DMS, where
possible.



The paper is structured as follows. In Section \ref{s:abinitio} the
\ai\ method used for generating DMS of CH\4\ is described and an
analytical representation for this DMS is introduced. The PES
is detailed in Section~\ref{s:PES}. The variational approach is described in Section~\ref{s:trove} and
theoretical vibrational transition moments and band intensities are reported in Section~\ref{s:tm}.
Finally, Section~\ref{s:concl} offers some discussion and future prospects.

\section{Ab initio calculations}
\label{s:abinitio}

We used the recently proposed explicitly correlated F12 coupled cluster method
with single and double excitations including a perturbational estimate of
connected triple excitations, CCSD(T)-F12c~\cite{07AdKnGe.method} in
conjunction with the corresponding F12-optimized correlation-consistent basis sets,
namely aug-cc-pVTZ-F12~\cite{08PeAdWe.method} (augmented correlation-consistent
polarized valence triple-zeta basis) as implemented in
Molpro 2010~\cite{molpro.method}.


Methane has nine vibrational degrees of freedom and thus requires a
very extensive grid of geometries in the \ai\ calculations. We have
computed three \ai\ DMSs (one for each Cartesian component) on a fine
grid of 114,000 geometries covering the energy range up to 50,000
\cm\ with C--H distances ranging from 0.75 to 2.0 \AA, and
interbond angles ranging from 50 to $140^\circ$.  The finite field
method, with a field of 0.005~a.u., was used to compute the dipole
moment component as first derivatives with respect to the external
electric field in the dipole approximation.

In order to represent the \ai\ dipole moment
vector $\vec\mu$ of CH\4\ analytically we use the so-called
symmetrized molecular bond (SMB) representation \cite{04MaQuxx.CH4},
where $\vec\mu$ is projected onto three symmetrized combinations of
vectors associated with the four molecular bonds C-H$_i$
($i=1,2,3,4$). An analogous approach was used to represent the \ai\
dipole moment of NH\3\ \cite{jt466} (see also Ref.~\cite{04MaQuxx.CH4}
for a non-symmetrized MB representation of the methane DMS). The three components of the
dipole moment of CH\4\ span three components of the $F_{2}$
representation, i.e. $F_{2x}$, $F_{2y}$, and $F_{2z}$ in the
irreducible representation of the \Td\ molecular symmetry
group \cite{98BuJexx.method}. We define the following three
symmetrically independent reference vectors $\vec n_{\Gamma}$ spanning
the same representation $F_2$:
\begin{eqnarray}
  \vec{n}_{F_{2x}} &=& \frac{1}{2} \left( \vecr{1} - \vecr{2} + \vecr{3} - \vecr{4} \right), \label{e:vects:x} \\
  \vec{n}_{F_{2y}} &=& \frac{1}{2} \left( \vecr{1} - \vecr{2} - \vecr{3} + \vecr{4} \right), \label{e:vects:y} \\
  \vec{n}_{F_{2z}} &=& \frac{1}{2} \left( \vecr{1} + \vecr{2} - \vecr{3} - \vecr{4} \right), \label{e:vects:z}
\end{eqnarray}
where $\vec{r}_{i}$ is a molecular bond vector pointing from C to H$_i$ ($i=1,2,3,4$).
The dipole moment vector $\vec\mu$ can now be represented as \begin{equation}\label{e:MB}
  \vec\mu = \vec{n}_{F_{2x}} \mu_{F_{2x}} + \vec{n}_{F_{2y}} \mu_{F_{2y}} + \vec{n}_{F_{2z}} \mu_{F_{2z}},
\end{equation}
where the projections $\mu_{F_{2\alpha}}$ ($\alpha= x,y,z$) are the dipole moment functions also spanning the $F_2$ representation.
Each component $\mu_{F_{2\alpha}}$ is then expanded as
\begin{equation}\label{e:DMS}
  \mu_{F_{2\alpha}} = \sum_{i_1,i_2,\ldots, i_9} \mu_{i_1,i_2,\ldots, i_9}^{\alpha} \xi_1^{i_1} \xi_2^{i_2} \ldots \xi_9^{i_9},
\end{equation}
in terms of the following set of nine independent internal coordinates:
\begin{eqnarray}
  \xi_i  &=& \left(r_i-r_{\rm e}\right) e^{-\beta (r_i-r_{\rm e})^2}, \;\; i = 1,2,3,4, \label{e:vects:i} \\
  \xi_5  &=& \frac{1}{\sqrt{12}} (2 \alpha_{12} - \alpha_{13} - \alpha_{14} - \alpha_{23} - \alpha_{24} + 2 \alpha_{34} ), \label{e:vects:5} \\
  \xi_6  &=& \frac{1}{2} (\alpha_{13} - \alpha_{14} - \alpha_{23} + \alpha_{24} ), \label{e:vects:6} \\
  \xi_7  &=& \frac{1}{\sqrt{2}} ( \alpha_{24} - \alpha_{13} ),  \label{e:vects:7} \\
  \xi_8  &=& \frac{1}{\sqrt{2}} ( \alpha_{23} - \alpha_{14} ), \label{e:vects:8}\\
  \xi_9  &=& \frac{1}{\sqrt{2}} ( \alpha_{34} - \alpha_{12} ), \label{e:vects:9}
\end{eqnarray}
where $r_i$ are the bond lengths and $\alpha_{ij}$ are the interbond
angles.  Following Hollenstein \textit{et al.}~\cite{94HoMaQu.CH4}, we
have introduced the factor $\exp\left(-\beta (r_i-r_{\rm e})^2\right)$
in order to keep the expansion in Eq.~(\ref{e:DMS}) from diverging at
large $r_i$.  Here $r_{\rm e}$ is a reference expansion center which
is conveniently taken at a value close to the molecular equilibrium.

The properties of the components $\mu_{F_{2\alpha}}$ associated with
$F_2$ impose symmetry relations between the expansion coefficients
$\mu_{i_1,\ldots,i_9}^{\alpha}$. These relations can be reconstructed
by applying the standard symmetry transformation rules
\cite{98BuJexx.method,11AlLeCa.CH4} between the $F_{2x},F_{2y},F_{2z}$
components of $F_2$. We employed the Maple program  function
`solve' to find a set of independent coefficients
$\mu_{i_1,\ldots,i_9}^{\alpha}$ by solving an over-determined system
of linear equations (see Ref.~\cite{05YuCaLi.NH3} for details on the
computational procedure). For example, to first order
the dipole moment functions $\mu_{\alpha}$ are given by the expansions
\begin{eqnarray}
\mu_{F_{2x}} &=& \mu_{7} \xi_7 +  \mu_{1} (\xi_1 - \xi_2 + \xi_3 - \xi_4) + O(\xi_k^2) \ldots, \label{e:mu:2nd:1} \\
\mu_{F_{2y}} &=& \mu_{7} \xi_8 +  \mu_{1} (\xi_1 - \xi_2 - \xi_3 + \xi_4) + O(\xi_k^2)\ldots, \label{e:mu:2nd:2}\\
\mu_{F_{2z}} &=& \mu_{7} \xi_9 +  \mu_{1} (\xi_1 + \xi_2 - \xi_3 - \xi_4) + O(\xi_k^2) \ldots, \label{e:mu:2nd:3}
\end{eqnarray}
where the shorthand notations $\mu_{7} = \mu_{0,0,0,0,0,0,1,0,0}^{\alpha}$ and $\mu_{1} = \mu_{1,0,0,0,0,0,0,0,0}^{\alpha}$ are used. The complete expansion
set is given in the supplementary material to this paper.
A full sixth-order symmetrized expansion requires 680 symmetrically
independent dipole moment parameters in total. Taking into account the
symmetry relations, the expansion Eq.~\eqref{e:DMS} can be written as:
\begin{equation}\label{e:mu-alpha}
  \mu_\alpha = \sum_{i_1,i_2,\ldots, i_9} \mu_{i_1,i_2,i_3,\ldots, i_9}^{\alpha}  \{\xi_1^{i_1} \xi_2^{i_2} \ldots \xi_9^{i_9} \}^{F_{2\alpha}},
\end{equation}
where $\{\xi_1^{i_1} \xi_2^{i_2} \ldots \xi_9^{i_9} \}^{F_{2\alpha}}$ is a symmetrized combination of different permutations of the internal coordinates consistent with \Td\ group operations and spanning  the $F_{2\alpha}$ symmetry.

The actual values of expansion parameters $\mu_{i_1,i_2,i_3,\ldots, i_9}^{\alpha}$ corresponding to our \ai\ DMS
were determined through least-squares fittings to the 114,000 \ai\
dipole moment values. Some parameters with large standard errors that
exhibit strong correlation were constrained to zero. The \ai\ data in these fits  were weighted
using the following expression proposed by \citet{ps97}:
\begin{equation}
\label{e:weights}
w_i = N \frac{\tanh \left[ -0.0004\;\mbox{cm} \times \left(V_i - 18\,000\;\mbox{cm}^{-1}\right)\right] + 1.00000002}
{ 2.00000002},
\end{equation}
where $N $ is a normalization factor defined by $\sum_i w_1 = 1$ and
$V_i$ is the
corresponding \ai\ energy at the $i$th geometry (in \cm), measured relative to the
equilibrium energy and computed using the same level of theory. The \ai\ energy $V_i$ is weighted by the factor
$w_i$ in the PES fitting; these weight factors favour the energies
below 18~000 cm$^{-1}$. In the end 296 parameters were needed to obtain a weighted ($w_i$) root-mean-squares (rms) error of
0.00016 D.  The expansion parameters are given in the supplementary materials of the paper along with a fortran 95 routine incorporating the corresponding analytical form.


\section{Potential energy surface}
\label{s:PES}

In this work we use a new `spectroscopic' PES generated as follows.
First, an \ai\ PES was generated on the same grid of 114,000
geometries employing the same CCSD(T)-F12c approach \cite{07AdKnGe.method} as for the DMS, but with
the larger aug-cc-pVQZ-F12 basis set \cite{08PeAdWe.method}. The frozen core approximation was applied
in the coupled cluster calculations.  We did not include core-valence corrections and higher-order coupled cluster corrections
because they cancel to a large extent (see the work by \citet{11YaYuRi.H2CS}),
and also because this would be very costly with the resources available to us.
Relativistic corrections were computed using the Douglas-Kroll method as
implemented in Molpro~2010.  With this method we obtained
the equilibrium bond length $r_{\rm e} = 1.08734$~\AA, which
is close to the value $r_{\rm e}$ = 1.08595~\AA\
derived in a combined experimental/\ai\ analysis~\cite{99Stxxxx.CH4}.
The \ai\ potential energies have been represented using a symmetrized
analytical expansion in terms of the four stretching coordinates
$1-\exp(-a (r_i-r_{\rm e}))$ ($i=1,2,3,4$) and five bending coordinates $\xi_{j}$
($j=5,6,7,8,9$) from Eqs.~(\ref{e:vects:i}--\ref{e:vects:9}). The
symmetry relations between potential parameters in the \Td\ group were
derived employing Maple with the procedure described above. In total
287 symmetrically independent parameters are needed for a full sixth-order
expansion. These expansion parameters were obtained from a fitting
procedure using the  weighting scheme given in Eq.~\eqref{e:weights}.
Some parameters showing large standard errors were constrained to
zero, which resulted in 268 potential parameters needed to get a weighted rms
error of 0.3~\cm.


To improve the quality of the \ai\ PES, we slightly refined
it by fitting to the `experimental' energies of $^{12}$CH\4\ extracted
from the  HITRAN~2008 \cite{HITRAN} database for $J=0,1,2,3,4$.
In these fittings the variational program TROVE \cite{07YuThJe.method}
was used, closely following the approach described in detail by
\citet{11YuBaTe.method}. This new `spectroscopic' PES for $^{12}$CH\4\
will be the subject of further improvements and a future publication. Both the initial \ai\ and
refined potential parameters are given as supplementary material along
with a fortran 95 program containing the corresponding potential
energy function.

\section{Variational calculations}
\label{s:trove}

The vibrational wavefunctions and energies of $^{12}$CH\4\ were calculated using the variational program TROVE \cite{07YuThJe.method}. In this approach the Hamiltonian operator is represented as an expansion around a reference geometry taken presently at the molecular equilibrium. The kinetic energy operator is expanded in terms of the nine linearized coordinates  $\xi_{i}\lnr$, which are linearized versions of the internal coordinates $\xi_i$ ($i=1,\ldots,9$) defined above; for the potential energy function we use $1-\exp(-a \xi_i\lnr)$ ($i=1,2,3,4$) for the four stretching and $\xi_j\lnr$ ($j=5,6,7,8,9$) for the five bending linearized coordinates. The latter expansions were applied to the refined PES introduced above. The kinetic and potential energy parts were truncated after the sixth and eighth order, respectively.

To construct the basis set, TROVE employs a multi-step contraction scheme based on the following polyad truncation. The polyad number in case of CH\4\ is given by
\begin{equation}\label{e:polyad}
  P = 2 (v_1 + v_2 + v_3 + v_4) + v_5 + v_6 + v_7 + v_8 + v_9,
\end{equation}
where $v_i$ is a vibrational quantum number associated with a
one-dimensional primitive basis function $\phi_{v_i}(\xi_{i}\lnr)$
($i=1,2,3,\ldots,9$).  At Step~1, nine sets of
$\phi_{v_i}(\xi_{i}\lnr)$ are generated using the Numerov-Cooley
\cite{61Coxxxx.method,24Nuxxxx.method} method by solving the 1D
\schr\ equations for each of the nine modes separately. The corresponding 1D
Hamiltonian operators are obtained from the 9D Hamiltonian operator
($J=0$) by freezing all but one vibrational coordinates at their
equilibrium values. At Step~2
the 9D coordinate space is divided into the three reduced sub-spaces,
(i) $\{\xi_1\lnr,\xi_2\lnr,\xi_3\lnr,\xi_4\lnr\}$, (ii)
$\xi_{5}\lnr,\xi_6\lnr$, and (iii) $\xi_7\lnr,\xi_8\lnr,\xi_9\lnr$.
This division is dictated by symmetry: each of the sub-spaces is
symmetrically independent and thus can be processed separately.  For
each of these sub-spaces, the \schr\ equations are solved for the
corresponding reduced Hamiltonian operators employing as basis the products of the
corresponding primitive functions $\phi_{v_i}(\xi_{i}\lnr)$ generated
at Step~1. The sub-space bases are truncated using the condition $P
\le P_{\rm max} $, where $P$ is given by Eq.~\eqref{e:polyad}. In this
work we use $P_{\rm max} = 12$, i.e.  our reduced sub-space basis sets
are defined by
\begin{eqnarray}
  v_1 + v_2 + v_3 + v_4 &\le& 6, \\
  v_5 + v_6 &\le& 12, \\
  v_7 + v_8 + v_9 &\le& 12.
\end{eqnarray}
The three sets of eigenfunctions
$\Psi_{\lambda_1}^{\rm (i)}(\xi_1\lnr,\xi_2\lnr,\xi_3\lnr,\xi_4\lnr)$,
$\Psi_{\lambda_2}^{\rm (ii)}(\xi_5\lnr,\xi_6\lnr$), and
$\Psi_{\lambda_3}^{\rm (iii)}(\xi_7\lnr,\xi_8\lnr,\xi_9\lnr)$ resulting from these
solutions are symmetrized and assigned local mode quantum numbers
(i) $v_1, v_2, v_3, v_4$, (ii) $v_5, v_6$, and (iii) $v_7,v_8,v_9$ and vibrational
symmetry $\Gamma_{\rm vib}$. The final vibrational basis set is formed
from products $\Psi_{\lambda_{1}}^{\rm (i)}\times \Psi_{\lambda_2}^{\rm (ii)} \times
\Psi_{\lambda_3}^{\rm (iii)}$, which are contracted with $P\le 12$ and
symmetrized again using the standard technique for transformations to
irreducible representations (see, for example, Ref.
\cite{98BuJexx.method}). Thus our contracted vibrational basis set
consists of ten symmetrically independent groups, one for each
irreducible representation of the \Td\ group $A_1$, $A_2$, $E_a$, $E_b$,
$F_{1x}$, $F_{1y}$, $F_{1z}$, $F_{2x}$, $F_{2y}$, $F_{2z}$. From the
degenerate symmetries $E, F_1$, and $F_2$, only the first component is
processed. Therefore only five vibrational Hamiltonian matrices (one
for each symmetry) are constructed and diagonalized. In the
diagonalizations we use the eigensolver DSYEV from LAPACK as
implemented in the MKL libraries. The dimensions of the matrices to be
diagonalized are 2190, 1725, 3901, 5481, and 5940 for $A_1$, $A_2$, $E$,
$F_1$, and $F_2$, respectively. The results are collected in
Tables~\ref{t:tm:gs}--\ref{t:tm:nu2}, where the calculated energies are
also compared with the available experimental values
from~\cite{98WeChxx.CH4,06BoReLo.CH4,09AlBaBo.CH4,10NiLyMi.CH4}.  It
should be noted that our basis set with $P \le 12$ does not provide
fully converged energies. The
same comment applies to our new `spectroscopic' PES: because of the
limitations of the method, this PES can only guarantee the energies
given in Table~\ref{t:tm:gs} when used with the same approach and
basis set employed in the refinements, i.e. $P_{\rm max}=12$. However the
vibrational transition moments reported in the next section appear to
be rather insensitive to the size of the basis set and the
shape of the potential.

 \section{Transition moments}
 \label{s:tm}

The transition moment connecting two states $i$ (`initial') and $f$ (`final') can be defined as the vibrational average of a dipole moment component $\mu_{\alpha}$ ($\alpha=x,y,z$):
\begin{equation}\label{e:mu:me}
\bar{\mu}^{(i,\Gamma_a\p;f,\Gamma_b\pp)}_{\alpha} =  \langle \Psi_{i}^{\Gamma_a\p} \vert \mu_{\alpha}  \vert \Psi_{f}^{\Gamma_b\pp} \rangle,
\end{equation}
where $\mu_{\alpha}$ ($\alpha=x,y,z$) are the Cartesian components of
the dipole moment in the molecule-fixed frame (defined in TROVE by the
Eckart conditions \cite{35Ecxxxx.method}); $\Psi_{i}^{\Gamma_a\p}$ and
$\Psi_{f}^{\Gamma_b\pp}$ are vibrational eigenfunctions obtained
variationally, and $\Gamma_s$ represents a component of one of the
ten irreducible representations $A_1$, $A_2$, $E_a$, $E_b$, $F_{1x}$,
$F_{1y}$, $F_{1z}$, $F_{2x}$, $F_{2y}$, $F_{2z}$ of \Td. It should be
noted that the transition moments defined by Eq.~\eqref{e:mu:me}
depend on the choice of the molecular-fixed coordinate system
\cite{jt121}  as
well as on the particular choice of the transformational properties of
the irreducible representations $E, F_1$, and $F_2$ (we use
irrep-matrices from Hougen's monograph \cite{01Hoxxxx.CH4}).  The
following quantity, averaged over the Cartesian components
$\alpha=x,y,z$ and any degenerate components of the wavefunctions, is
independent of these choices:
\begin{equation}\label{e:tm}
   \bar{\mu}^{if} \equiv  \bar{\mu}^{(i,\Gamma\p;f,\Gamma\pp)} = \sqrt{
    \left[\bar{\mu}_x^{(i,\Gamma\p;f,\Gamma\pp)}\right]^2+
    \left[\bar{\mu}_y^{(i,\Gamma\p;f,\Gamma\pp)}\right]^2+
    \left[\bar{\mu}_z^{(i,\Gamma\p;f,\Gamma\pp)}\right]^2  },
\end{equation}
where
\begin{equation}\label{e:tm:ab}
   \bar{\mu}_\alpha^{(i,\Gamma\p;f,\Gamma\pp)} = \sum_{a,b} \bar{\mu}_\alpha^{(i,\Gamma_a\p;f,\Gamma_b\pp)} ,
\end{equation}
and the vibrational transition moment $\bar{\mu}^{if}$ is defined as the length of the vector \big[$\bar{\mu}_x^{(i,\Gamma\p;f,\Gamma\pp)}$, $\bar{\mu}_y^{(i,\Gamma\p;f,\Gamma\pp)}$, $\bar{\mu}_z^{(i,\Gamma\p;f,\Gamma\pp)}$\big].


A number  of calculated values of the vibrational transition moments  $\bar{\mu}^{if}$ as well as matrix elements  $\bar{\mu}_x^{(i,\Gamma\p;f,\Gamma\pp)}$ of $^{12}$CH\4\ corresponding to transitions from the three lowest vibrational states ($\nu_0$, $\nu_2$, and $\nu_4$) are listed in Tables~\ref{t:tm:gs}--\ref{t:tm:nu2}. The non-zero matrix elements $\bar{\mu}_x^{(i,\Gamma\p;f,\Gamma\pp)}$ in Tables~\ref{t:tm:gs}--\ref{t:tm:nu2} for each transition satisfy the following set of relations:
\begin{equation}
\label{e:mu:rel}
\begin{array}{c@{}c@{}c@{}c@{}c@{}c@{}c@{}c@{}c@{}c@{}c}
\tabcolsep=0pt
  \bar{\mu}^{(A_1;F_{2z})}_{x} &=& -\bar{\mu}^{(A_1;F_{2x})}_{y} &=& -\bar{\mu}^{(A_1;F_{2y})}_{z}, \\
  \bar{\mu}^{(F_{2x};F_{2y})}_{x} &=&  - \bar{\mu}^{(F_{2y};F_{2z})}_{y} &=&  -\bar{\mu}^{(F_{2x};F_{2z})}_{z},   \\
  \bar{\mu}^{(F_{2z};E_{a})}_{x} &=&
  2\bar{\mu}^{(F_{2x};E_{a})}_{y} &=&   -\frac{2}{\sqrt{3}} \bar{\mu}^{(F_{2x};E_{b})}_{y} &=&
  2\bar{\mu}^{(F_{2y};E_{a})}_{z} &=&   \frac{2}{\sqrt{3}} \bar{\mu}^{(F_{2z};E_{b})}_{z},   \\
  \bar{\mu}^{(F_{2x};F_{1z})}_{x} &=&  - \bar{\mu}^{(F_{2y};F_{1y})}_{x} &=&  -\bar{\mu}^{(F_{2y};F_{1x})}_{y}  &=&
  \bar{\mu}^{(F_{2y};F_{1z})}_{y} &=&   \bar{\mu}^{(F_{2z};F_{1x})}_{z}  &=&  -\bar{\mu}^{(F_{2z};F_{1y})}_{z}, \\
  \bar{\mu}^{(F_{1x};E_{b})}_{x} &=&
  \frac{2}{\sqrt{3}}\bar{\mu}^{(F_{1y};E_{a})}_{y} &=&   2 \bar{\mu}^{(F_{1y};E_{b})}_{y} &=&
  -\frac{2}{\sqrt{3}}\bar{\mu}^{(F_{1z};E_{a})}_{z} &=&   2 \bar{\mu}^{(F_{1z};E_{b})}_{z},
\end{array}
\end{equation}
where indices $i,f$ are omitted for simplicity. Therefore  Tables~\ref{t:tm:gs}--\ref{t:tm:nu2} list only one value for each pair of states.
A complete list of 47861 vibrational transition moments and matrix elements $\bar{\mu}_\alpha^{(i,\Gamma_a\p;f,\Gamma_b\pp)}$ of $^{12}$CH\4\ is provided as supplementary material for all allowed transitions in the frequency range from 0 to 10000 \cm\ with lower state energies below 5000~\cm.


Because of the high symmetry of methane many vibrational transition moments
$\bar{\mu}^{if}$ vanish. For example,
$\bar{\mu}^{if}\equiv0$ between $A_1$ eigenfunctions of the ground
vibrational state (GS) and any vibrational states of $A_1$, $A_2$, $E$,
and $F_1$ symmetry. This includes transitions within the GS and
the transitions giving rise to the fundamental bands $\nu_1 (A_1)$ and $\nu_2 (E)$.
All these transitions have zero vibrational transition moments and are thus dipole forbidden.
Certainly this does not rule out ro-vibrational transitions within these bands
although such transitions are often also referred to as `forbidden'. At
a formal level any transitions connecting rotation-vibration states
with the correct symmetries are allowed \cite{jt72,jt89}. In practice,
centrifugal distortions arising from rotation-vibration
interactions can introduce $F_2$-vibrational contributions into, for
example, the GS rotational wavefunctions, thus
producing non-zero ro-vibrational line strengths. Conversely,
`allowed' ro-vibrational transitions are based only on non-zero
vibrational transition moments
$\bar{\mu}^{(i,\Gamma_a\p;f,\Gamma_b\pp)}_{\alpha}$, such as the ones
given in Tables~\ref{t:tm:gs}--\ref{t:tm:nu2}.

The transition moments presented here
can be used to derive the effective dipole moment parameters of
$^{12}$CH\4, which are important for modelling intensities of this
molecule (see Ref.~\cite{09AlBaBo.CH4} for the most
recent collection of the effective dipole moment parameters). Examples
of such an exercise can be found in
Refs.~\cite{91PeLyTy.CH4,98MoAbLo.CH4}. For example, the
parameter $V^{0(0, 0A1)}_{0000A_1-0001 F_2} = 0.098677(88)$~D
\cite{09AlBaBo.CH4} is in good agreement with our value $\sqrt{3}
\bar{\mu}^{({\rm g.s.},A_1;\nu_4,F_2)}_{x}$  =
0.098314~D. Similarly, $V^{0(0, 0A1)}_{0000 A_1-0010 F_2} =
−0.094622(47)$~D \cite{09AlBaBo.CH4} agrees well with our value
$\sqrt{3} \bar{\mu}^{({\rm g.s.},A_1;\nu_3,F_2)}_{x}$ =
0.093699~D (see Table~\ref{t:tm:gs}). Here $\sqrt{3}$ is a conversion factor
associated with the Dyad-GS and Pentad-GS zero-order effective dipole
constants in the tensorial representation (see, for example,
Refs.~\cite{89BrLoHi.CH4,92HiLoBr.CH4}). For higher vibrational order parameters
the correspondence between our transition moments and effective dipole
moment parameters is less straightforward owing to the complexity of
the effective dipole moment operators which are usually given in a tensorial
representation.




From the transition moments computed we also generated vibrational band intensities as given by \citet{89BrLoHi.CH4} and \cite{92HiLoBr.CH4}:
\begin{equation}\label{e:band-int}
  S_{\rm vib} = \frac{8 \pi^3 10^{-36}}{3 h c Q_{v}}\frac{L T_0}{T} e^{-E_{\rm i}/kT}   \nu_{if} \left( 1 - e^{-hc \nu_{if}/kT} \right) \bar{\mu}_{if}^2,
\end{equation}
where $\nu_{if}$ and $E_{\rm i}$ are the vibrational transition
frequency and the lower state energy, respectively, Lochschmidt's number
$L=2.686754\times
10^{19}$ cm$^{-3}$, $T_0$ = 273.15~K, $k$ is the Boltzmann constant, and
$Q_v$ = 1.002 is the vibrational partition function for $T=278.15$~K.
Some of the computed values of $ S_{\rm vib}$ are also illustrated in
Tables~\ref{t:tm:gs}--\ref{t:tm:nu2}, where they are compared to the
experimental band intensities from
Refs.~\cite{92HiLoBr.CH4,92HiLoBr.CH4,96OuHiLo.CH4,01HiRoLo.CH4,01RoHiLo.CH4,09AlBaBo.CH4}, where available. The agreement
is generally very good. In some cases, particularly for combinational bands,
there is some ambiguity over how to distribute the individual lines
between transitions with the same normal-mode vibrational quantum
numbers (see also footnotes to Tables~\ref{t:tm:gs}--\ref{t:tm:nu2}).
This introduces an extra uncertainty into these comparisons.

\begin{figure}[t]
\begin{center}
{\leavevmode \epsfxsize=11.0cm \epsfbox{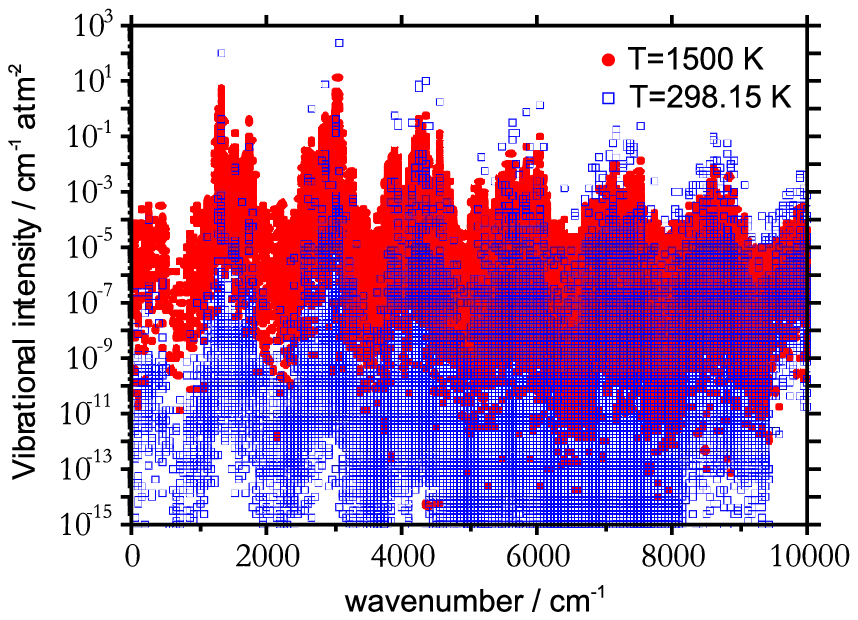}}
\end{center}
\caption{\label{f:S:300:1500}  Vibrational band intensities calculated at $T=$ 298.15~K (filled blue circles) and $T=$ 1500~K (empty red squares) using Eq.~\eqref{e:band-int}.
}
\end{figure}

As an illustration of a possible application of our transition moments, Fig.~\ref{f:S:300:1500} shows the vibrational band intensities calculated at $T=$ 298.15~K (empty blue squares) and $T=$ 1500~K (filled red circles) using Eq.~\eqref{e:band-int} ($Q_v$ = 3.038 at $T=1500$~K).


\section{Conclusion}
\label{s:concl}

In this work we present a new \ai\ DMS for CH\4\ which is used to
generate the vibrational transition moments and band intensities of
$^{12}$CH\4\ for a large number of vibrationally allowed transitions.
These values should be helpful for predicting intensities for individual
bands. The dipole matrix elements can also be used for
predicting or evaluating effective dipole moment parameters.  The analytical representations
for the electric dipole moment and potential energy functions of CH\4\ developed in this work
have correct symmetry properties of the \Td\ group and should be useful for representing similar objects
obtained using different methods or levels of theory.


Comparisons of calculated and experimental (or experimentally
derived) band intensities demonstrate the good quality of our new DMS of
CH\4.  We are planning to use this DMS in calculations of a high
temperature line list for methane within the framework of the ExoMol
project \cite{jt528} (see {\it{www.exomol.com}}), which has
identified methane as a key target species. Such a line list
will be important for modelling molecular opacity in atmospheres of
(exo-)planets and cool stars.


\begin{table}
\scriptsize
\caption{\label{t:tm:gs}
Vibrational transition moments, $\bar\mu^{if}$ in D, individual matrix elements $\bar{\mu}^{(i,A_1;f,F_{2z})}_{x}$ (D),  and band intensities, $S_{if}$ in cm$^{-1}$atm$^{-2}$, for $^{12}$CH\4\ for transitions from the vibrationally ground state. The calculated and experimentally derived term values $\bar{E}_{\rm f}$ of the upper states are given in \cm.}
\begin{center}
\tabcolsep=5pt
\vspace*{5pt}
\begin{tabular}{lcccrcrr@{}l}
    \hline
    \hline
$\Gamma$ & State-f & $\bar{E}_{\rm f}^{\rm obs}$  & $\bar{E}_{\rm f}^{\rm calc}$ &  $\bar{\mu}^{(i,A_1;f,F_{2z})}_{x}$  & $\bar\mu^{if}_{\rm calc}$  & $S_{if}^{\rm(calc)}$ & $S_{if}^{\rm(obs)}$\cite{09AlBaBo.CH4} \\
 \hline
$F_2$     &  0 0 0 1   &   1310.76  &      1310.87  &       -0.05676  &   0.09831  &        129.323  &  127.68      &                                \\
$F_2$     &  0 0 0 2   &   2614.26  &      2614.32  &        0.00408  &   0.00706  &          1.334  &  1.05        &                                \\
$F_2$     &  0 1 0 1   &   2830.32  &      2830.28  &        0.01038  &   0.01797  &          9.346  &  6.63        &$^a$                            \\
$F_2$     &  0 0 1 0   &   3019.49  &      3019.49  &       -0.05410  &   0.09370  &        271.062  &  269.92      &                                \\
$F_2$     &  0 0 0 3   &   3870.49  &      3870.49  &       -0.00231  &   0.00401  &          0.636  &  0.59        &  \multirow{2}{*}{$\Big\}b$}    \\
$F_2$     &  0 0 0 3   &   3930.92  &      3930.81  &        0.00166  &   0.00288  &          0.332  &  0.14        &                                \\
$F_2$     &  0 1 0 2   &   4142.86  &      4142.99  &       -0.00171  &   0.00295  &          0.370  &  0.34        &                                \\
$F_2$     &  1 0 0 1   &   4223.46  &      4223.52  &        0.00775  &   0.01343  &          7.791  &  7.84        &                                \\
$F_2$     &  0 0 1 1   &   4319.21  &      4319.43  &       -0.00919  &   0.01593  &         11.201  &  10.24       &                                \\
$F_2$     &  0 2 0 1   &   4348.72  &      4348.97  &       -0.00145  &   0.00252  &          0.281  &  0.55        &  \multirow{2}{*}{$\Big\}c$}    \\
$F_2$     &  0 2 0 1   &   4378.95  &      4379.15  &       -0.00013  &   0.00023  &          0.002  &  0.23        &                                \\
$F_2$     &  0 1 1 0   &   4543.76  &      4543.90  &        0.00392  &   0.00678  &          2.138  &  1.26        &$^d$                            \\
$F_2$     &  0 0 0 4   &   5143.24  &      5143.25  &        0.00035  &   0.00060  &          0.019  &  0.01        &$^e$                            \\
$F_2$     &  0 0 0 4   &   5211.29  &      5210.65  &       -0.00021  &   0.00037  &          0.007  &              &                                \\
$F_2$     &  0 1 0 3   &   5370.52  &      5370.37  &       -0.00014  &   0.00023  &          0.003  &              &                                \\
$F_2$     &  0 1 0 3   &   5429.58  &      5429.07  &       -0.00058  &   0.00101  &          0.056  &              &                                \\
$F_2$     &  0 1 0 3   &   5445.12  &      5444.87  &        0.00011  &   0.00020  &          0.002  &              &                                \\
$F_2$     &  1 0 0 2   &            &      5519.47  &        0.00011  &   0.00019  &          0.002  &              &                                \\
$F_2$     &  0 0 1 2   &   5588.03  &      5586.76  &        0.00156  &   0.00270  &          0.416  &              &                                \\
$F_2$     &  0 0 1 2   &            &      5620.63  &       -0.00074  &   0.00129  &          0.095  &              &                                \\
$F_2$     &  0 0 1 2   &            &      5633.69  &       -0.00102  &   0.00176  &          0.179  &              &                                \\
$F_2$     &  0 2 0 2   &   5643.45  &      5643.64  &        0.00005  &   0.00009  &          0.000  &              &                                \\
$F_2$     &  0 2 0 2   &   5668.60  &      5668.70  &       -0.00027  &   0.00047  &          0.013  &              &                                \\
$F_2$     &  1 1 0 1   &            &      5727.16  &        0.00016  &   0.00028  &          0.005  &              &                                \\
$F_2$     &  0 1 1 1   &   5823.10  &      5823.03  &       -0.00213  &   0.00368  &          0.808  &              &                                \\
$F_2$     &  0 1 1 1   &   5844.00  &      5843.72  &       -0.00052  &   0.00089  &          0.048  &              &                                \\
$F_2$     &  1 0 1 0   &            &      5861.92  &       -0.00031  &   0.00054  &          0.018  &              &                                \\
$F_2$     &  0 3 0 1   &   5867.66  &      5868.23  &       -0.00026  &   0.00046  &          0.012  &              &                                \\
$F_2$     &  0 3 0 1   &   5894.12  &      5894.54  &       -0.00007  &   0.00013  &          0.001  &              &                                \\
$F_2$     &  0 0 2 0   &   6004.69  &      6004.45  &       -0.00301  &   0.00522  &          1.671  &  1.63        &$^e$                                 \\
$F_2$     &  0 2 1 0   &   6054.64  &      6054.67  &       -0.00055  &   0.00094  &          0.055  &              &                                \\
$F_2$     &  0 2 1 0   &   6065.32  &      6065.26  &       -0.00055  &   0.00095  &          0.056  &              &                                \\
\hline
\hline
\end{tabular}
\end{center}
\vspace*{5pt}

$^a$ The total intensity of the $\nu_2+\nu_4$ band, 9.94 cm$^{-1}$atm$^{-2}$~\citet{09AlBaBo.CH4}, which includes both $F_1$  and  $F_2$ sub-components agrees well with
our $S^{\rm (calc)}_{0101F_2}$ value.

\noindent
$^b$ The total intensity of the $3\nu_4$ band, 0.83 cm$^{-1}$atm$^{-2}$~\citet{09AlBaBo.CH4}, which includes  $A_1$ and $F_1$ sub-components in addition to two $F_2$ ones can be compared
to the sum of the two `calc' band intensities $S^{\rm (calc)}_{0003F_2}$.

\noindent
$^c$ The total intensity of the $2\nu_2+\nu_4$ band which includes three sub-bands ($F_1$, $F_2$ and $F_2$) by \citet{09AlBaBo.CH4} is 0.93 cm$^{-1}$atm$^{-2}$.
\citet{01HiRoLo.CH4} reported somewhat smaller value of 0.63 cm$^{-1}$atm$^{-2}$, which suggests a better agreement with our value for the $F_2$ sub-band only.

\noindent
$^d$ Compare to the estimate by  \citet{92HiLoBr.CH4}, 1.84 cm$^{-1}$atm$^{-2}$, which is closer to our value.

\noindent
$^e$ As estimated by \citet{01RoHiLo.CH4}.

\end{table}

\begin{table}
\scriptsize
\caption{\label{t:tm:nu4}
Vibrational transition moments, $\bar\mu^{if}$ in D, individual matrix elements $\bar{\mu}_x^{(i,\Gamma_{\alpha};f,F_{2\beta})}$ ($\alpha=x,y,z$ or $a,b$ and
$\beta=x,y,z$)  in D, and band intensities, $S_{if}$ in cm$^{-1}$atm$^{-2}$, for $^{12}$CH\4\ for the transitions from the $\nu_4$ state.
The calculated and experimentally derived term values $\bar{E}_{\rm f}$ of the upper states are given in \cm. }
\begin{center}
\vspace*{5pt}
\begin{tabular}{lcccccrcrr}
    \hline
    \hline
$\Gamma_{\rm f}$ & State-f & $\bar{E}_{\rm f}^{\rm obs}$ & $\bar{E}_{\rm f}^{\rm calc}$  &  $\bar{\mu}_x^{(i,\Gamma_{\alpha};f,F_{2\beta})}$  & $\alpha$ & $\beta$ &
$\bar\mu^{if}_{\rm calc}$   & $S_{if}^{\rm (calc)}$ & $S_{if}^{\rm (obs)}$\cite{96OuHiLo.CH4} \\
 \hline
$E$       &  0 1 0 0   &   1533.33  &      1533.39  &       -0.00406 &$   a   $&$     z   $&     0.00704  &        0.00013  &                                          \\
$A_1$     &  0 0 0 2   &   2587.04  &      2586.98  &        0.05046 &$       $&$     z   $&     0.08741  &        0.17802  &  \multirow{3}{*}{$\Bigg\}$0.931$^a$}     \\
$F_2$     &  0 0 0 2   &   2614.26  &      2614.32  &        0.05718 &$   y   $&$     x   $&     0.14006  &        0.46701  &                                          \\
$E$       &  0 0 0 2   &   2624.62  &      2624.61  &        0.06530 &$   a   $&$     z   $&     0.11310  &        0.30697  &                                          \\
$F_2$     &  0 1 0 1   &   2830.32  &      2830.28  &        0.00641 &$   y   $&$     x   $&     0.01569  &        0.00684  &  0.00872                                 \\
$F_1$     &  0 1 0 1   &   2846.07  &      2846.06  &        0.00007 &$   z   $&$     x   $&     0.00016  &       0.000001  &                                          \\
$A_1$     &  1 0 0 0   &   2916.48  &      2916.54  &        0.00236 &$       $&$     z   $&     0.00408  &        0.00049  &  0.000689                                \\
$F_2$     &  0 0 1 0   &   3019.49  &      3019.49  &       -0.01675 &$   a   $&$     z   $&     0.04104  &        0.05265  &  0.0481                                  \\
$A_1$     &  0 2 0 0   &   3063.65  &      3063.61  &       -0.00088 &$       $&$     z   $&     0.00153  &        0.00007  &  0.0005                                  \\
$E$       &  0 2 0 0   &   3065.14  &      3065.12  &        0.00144 &$   a   $&$     z   $&     0.00250  &        0.00020  &                                          \\
$F_2$     &  0 0 0 3   &   3870.49  &      3870.49  &       -0.00383 &$   a   $&$     z   $&     0.00938  &        0.00412  &                                          \\
$A_1$     &  0 0 0 3   &   3909.20  &      3909.68  &       -0.00424 &$       $&$     z   $&     0.00734  &        0.00256  &                                          \\
$F_1$     &  0 0 0 3   &   3920.51  &      3920.15  &       -0.00320 &$   z   $&$     x   $&     0.00784  &        0.00293  &                                          \\
$F_2$     &  0 0 0 3   &   3930.92  &      3930.81  &       -0.00299 &$   a   $&$     z   $&     0.00732  &        0.00257  &                                          \\
$E$       &  0 1 0 2   &   4101.39  &      4101.49  &       -0.00973 &$   a   $&$     z   $&     0.01685  &        0.01451  &                                          \\
$F_1$     &  0 1 0 2   &   4128.76  &      4128.73  &        0.00890 &$   z   $&$     x   $&     0.02180  &        0.02451  &                                          \\
$A_1$     &  0 1 0 2   &   4132.86  &      4132.83  &        0.00923 &$       $&$     z   $&     0.01599  &        0.01320  &                                          \\
$F_2$     &  0 1 0 2   &   4142.86  &      4142.99  &       -0.00537 &$   a   $&$     z   $&     0.01315  &        0.00897  &                                          \\
$E$       &  0 1 0 2   &   4151.21  &      4151.28  &        0.00525 &$   a   $&$     z   $&     0.00910  &        0.00430  &                                          \\
$F_2$     &  1 0 0 1   &   4223.46  &      4223.52  &        0.00244 &$   a   $&$     z   $&     0.00597  &        0.00190  &                                          \\
$F_2$     &  0 0 1 1   &   4319.21  &      4319.43  &       -0.03798 &$   a   $&$     z   $&     0.09303  &        0.47649  &                                          \\
$E$       &  0 0 1 1   &   4322.18  &      4322.21  &       -0.04272 &$   a   $&$     z   $&     0.07399  &        0.30170  &                                          \\
$A_1$     &  0 0 1 1   &   4322.70  &      4322.40  &       -0.03326 &$       $&$     z   $&     0.05761  &        0.18292  &                                          \\
$F_1$     &  0 0 1 1   &   4322.59  &      4322.66  &       -0.03760 &$   z   $&$     x   $&     0.09211  &        0.46758  &                                          \\
$F_2$     &  0 2 0 1   &   4348.72  &      4348.97  &       -0.00451 &$   a   $&$     z   $&     0.01104  &        0.00678  &                                          \\
$F_1$     &  0 2 0 1   &   4363.61  &      4363.61  &        0.00076 &$   z   $&$     x   $&     0.00186  &        0.00019  &                                          \\
$F_2$     &  0 2 0 1   &   4378.95  &      4379.15  &       -0.00057 &$   a   $&$     z   $&     0.00140  &        0.00011  &                                          \\
$E$       &  1 1 0 0   &   4435.12  &      4435.18  &       -0.00095 &$   a   $&$     z   $&     0.00165  &        0.00015  &                                          \\
$F_1$     &  0 1 1 0   &   4537.55  &      4537.75  &        0.00143 &$   z   $&$     x   $&     0.00351  &        0.00073  &                                          \\
$F_2$     &  0 1 1 0   &   4543.76  &      4543.90  &       -0.00121 &$   a   $&$     z   $&     0.00297  &        0.00052  &                                          \\
$E$       &  0 3 0 0   &   4592.03  &      4592.12  &       -0.00030 &$   a   $&$     z   $&     0.00052  &        0.00002  &                                          \\
$A_1$     &  0 3 0 0   &   4595.51  &      4595.66  &        0.00000 &$       $&$     z   $&     0.00000  &        0.00000  &                                          \\
\hline
\hline
\end{tabular}
\end{center}
\noindent $^a$  This `experimental' vibrational intensity of the $2\nu_4$ band \cite{96OuHiLo.CH4} correlates with our $S^{\rm (calc)}_{0002A_1}$ + $S^{\rm (calc)}_{0002F_2}$ + $S^{\rm (calc)}_{0002E}$  = 0.952 cm$^{-1}$atm$^{-2}$.

\end{table}

\begin{table}
\scriptsize
\caption{\label{t:tm:nu2}
Vibrational transition moments, $\bar\mu^{if}$ in D, individual matrix elements $\bar{\mu}_x^{(i,\Gamma_{\alpha};f,E_{\beta})}$ ($\alpha=x,y,z$ and $\beta=a,b$) in D. and band intensities, $S_{if}$ in cm$^{-1}$atm$^{-2}$, for
$^{12}$CH\4\ for the transitions from the $\nu_2$ state. The calculated and experimentally derived term values $\bar{E}_{\rm f}$ of the upper states are given in \cm.
 }
\begin{center}
\vspace*{5pt}
\begin{tabular}{lcccccrcrr}
    \hline
    \hline
$\Gamma_{\rm f}$ & State-f & $\bar{E}_{\rm f}^{\rm obs}$ & $\bar{E}_{\rm f}^{\rm calc}$  &  $\bar{\mu}_x^{(i,\Gamma_{\alpha};f,E_{\beta})}$  & $\alpha$ & $\beta$ &
$\bar\mu^{if}_{\rm calc}$  & $S_{if}^{\rm (calc)}$ & $S_{if}^{\rm (obs)}$\cite{96OuHiLo.CH4} \\
 \hline
$F_2$     &  0 0 0 2   &   2614.26  &      2614.32  &        0.00110 &$    z   $&$   a    $&     0.00191  &        0.00002  &  0.00008                                 \\
$F_2$     &  0 1 0 1   &   2830.32  &      2830.28  &       -0.05740 &$    z   $&$   a    $&     0.09942  &        0.08000  &  \multirow{2}{*}{$\Bigg\}$0.145$^a$}     \\
$F_1$     &  0 1 0 1   &   2846.07  &      2846.06  &       -0.05718 &$    x   $&$   b    $&     0.09904  &        0.08037  &                                          \\
$F_2$     &  0 0 1 0   &   3019.49  &      3019.49  &       -0.01835 &$    z   $&$   a    $&     0.03178  &        0.00938  &  0.0085                                  \\
$F_2$     &  0 0 0 3   &   3870.49  &      3870.49  &        0.00030 &$    z   $&$   a    $&     0.00052  &       0.000004  &                                          \\
$F_1$     &  0 0 0 3   &   3920.51  &      3920.15  &       -0.00034 &$    x   $&$   b    $&     0.00058  &       0.000005  &                                          \\
$F_2$     &  0 0 0 3   &   3930.92  &      3930.81  &       -0.00010 &$    z   $&$   a    $&     0.00017  &      0.0000005  &                                          \\
$F_1$     &  0 1 0 2   &   4128.76  &      4128.73  &       -0.00367 &$    x   $&$   b    $&     0.00636  &        0.00066  &                                          \\
$F_2$     &  0 1 0 2   &   4142.86  &      4142.99  &       -0.00292 &$    z   $&$   a    $&     0.00507  &        0.00042  &                                          \\
$F_2$     &  1 0 0 1   &   4223.46  &      4223.52  &       -0.00310 &$    z   $&$   a    $&     0.00536  &        0.00048  &                                          \\
$F_2$     &  0 0 1 1   &   4319.21  &      4319.43  &       -0.00142 &$    z   $&$   a    $&     0.00245  &        0.00010  &                                          \\
$F_1$     &  0 0 1 1   &   4322.59  &      4322.66  &       -0.00218 &$    x   $&$   b    $&     0.00377  &        0.00025  &                                          \\
$F_2$     &  0 2 0 1   &   4348.72  &      4348.97  &        0.01360 &$    z   $&$   a    $&     0.02355  &        0.00977  &                                          \\
$F_1$     &  0 2 0 1   &   4363.61  &      4363.61  &       -0.01117 &$    x   $&$   b    $&     0.01934  &        0.00662  &                                          \\
$F_2$     &  0 2 0 1   &   4378.95  &      4379.15  &       -0.00081 &$    z   $&$   a    $&     0.00139  &        0.00003  &                                          \\
$F_1$     &  0 1 1 0   &   4537.55  &      4537.75  &       -0.05404 &$    x   $&$   b    $&     0.09360  &        0.16459  &                                          \\
$F_2$     &  0 1 1 0   &   4543.76  &      4543.90  &        0.05344 &$    z   $&$   a    $&     0.09255  &        0.16125  &                                          \\
$F_2$     &  0 0 0 4   &   5143.24  &      5143.25  &       -0.00007 &$    z   $&$   a    $&     0.00011  &        0.00000  &                                          \\
$F_2$     &  0 0 0 4   &   5211.29  &      5210.65  &       -0.00011 &$    z   $&$   a    $&     0.00019  &        0.00000  &                                          \\
$F_1$     &  0 0 0 4   &   5230.78  &      5230.07  &        0.00003 &$    x   $&$   b    $&     0.00005  &        0.00000  &                                          \\
$F_2$     &  0 1 0 3   &   5370.52  &      5370.37  &       -0.00262 &$    z   $&$   a    $&     0.00454  &        0.00049  &                                          \\
$F_1$     &  0 1 0 3   &   5389.67  &      5389.71  &        0.00191 &$    x   $&$   b    $&     0.00331  &        0.00026  &                                          \\
$F_2$     &  0 1 0 3   &   5429.58  &      5429.07  &        0.00117 &$    z   $&$   a    $&     0.00202  &        0.00010  &                                          \\
$F_1$     &  0 1 0 3   &   5436.79  &      5436.50  &        0.00102 &$    x   $&$   b    $&     0.00176  &        0.00008  &                                          \\
$F_2$     &  0 1 0 3   &   5445.12  &      5444.87  &       -0.00064 &$    z   $&$   a    $&     0.00110  &        0.00003  &                                          \\
$F_1$     &  0 1 0 3   &   5462.92  &      5462.74  &        0.00198 &$    x   $&$   b    $&     0.00343  &        0.00029  &                                          \\
\hline
\hline
\end{tabular}
\end{center}
\vspace*{5pt}


\noindent $^a$  This `experimental' vibrational intensity of the $\nu_2+\nu_4$ band \cite{96OuHiLo.CH4} correlates with our $S^{\rm (calc)}_{0101F_1}$ + $S^{\rm (calc)}_{0101F_2}$ = 0.160 cm$^{-1}$atm$^{-2}$.




\end{table}

\vspace{1.5in}

\section*{Acknowledgments}
This work is supported by ERC Advanced Investigator Project 267219 and
by the UK Science and Facilities Research Council (STFC).

\vspace{1.5in}

\bibliographystyle{model1a-num-names}

\end{document}